\documentclass[11pt,twoside,a4paper,cr,final]{cms-tdr}
\def\svnVersion{77442M}

\begin{document}

\hyphenation{had-ron-i-za-tion}
\hyphenation{cal-or-i-me-ter}
\hyphenation{de-vices}
\RCS$Revision: 74250 $
\RCS$HeadURL: svn+ssh://svn.cern.ch/reps/tdr2/papers/HIN-10-006/trunk/QM2011_QuarkoniaCMSPlenary_Silvestre.tex $
\RCS$Id: QM2011_QuarkoniaCMSPlenary_Silvestre.tex 74250 2011-08-10 18:56:34Z silvest $

\newcommand{\ee}{\ensuremath{e^+e^-}\xspace}
\newcommand{\mumu}{\ensuremath{\mu^+\mu^-}\xspace}
\newcommand{\eexp}[1]{\ensuremath{{\text e}^{#1}}\xspace}

\newcommand{\qqbar}{\ensuremath{{\cmsSymbolFace{q}\overline{\cmsSymbolFace{q}}}}\xspace}
\newcommand{\QQbar}{\ensuremath{{\cmsSymbolFace{Q}\overline{\cmsSymbolFace{Q}}}}\xspace}
\newcommand{\Jpsi}{\ensuremath{\cmsSymbolFace{J}\hspace{-.08em}/\hspace{-.14em}\psi}\xspace} 
\newcommand{\B}{\ensuremath{\cmsSymbolFace{B}}\xspace}
\renewcommand{\PgU}{\ensuremath{\Upsilon}\xspace}
\renewcommand{\PgUa}{\ensuremath{\Upsilon\text{(1S)}}\xspace}
\renewcommand{\PgUb}{\ensuremath{\Upsilon\text{(2S)}}\xspace}
\renewcommand{\PgUc}{\ensuremath{\Upsilon\mathrm{(3S)}}\xspace}
\newcommand{\PgUbc}{\ensuremath{\Upsilon\text{(2S+3S)}}\xspace}
\newcommand{\PgUn}{\ensuremath{\Upsilon\text{(nS)}}\xspace}
\newcommand{\z}{\ensuremath{\cmsSymbolFace{Z}}\xspace}

\newcommand{\dndy}{\ensuremath{dN/dy}\xspace}
\newcommand{\dnchdy}{\ensuremath{dN_{\text{ch}}/dy}\xspace}
\newcommand{\dndeta}{\ensuremath{dN/d\eta}\xspace}
\newcommand{\dnchdeta}{\ensuremath{dN_{\text{ch}}/d\eta}\xspace}
\newcommand{\dndpt}{\ensuremath{dN/d\pt}\xspace}
\newcommand{\dnchdpt}{\ensuremath{dN_{\text{ch}}/d\pt}\xspace}
\newcommand{\deta}{\ensuremath{\Delta\eta}\xspace}
\newcommand{\dphi}{\ensuremath{\Delta\phi}\xspace}

\newcommand {\npart}  {\ensuremath{N_{\text{part}}}\xspace}
\newcommand {\ncoll}  {\ensuremath{N_{\text{coll}}}\xspace}

\newcommand{\raa}{\ensuremath{R_{AA}}\xspace}
\newcommand{\taa}{\ensuremath{T_{AA}}\xspace}

\newcommand{\eq}[1]{Eq.~\eqref{#1}\xspace}
\newcommand{\fig}[1]{Fig.~\ref{#1}\xspace}
\newcommand{\tab}[1]{Tab.~\ref{#1}\xspace}

\newcommand{\pp}{{\ensuremath{\Pp\Pp}}\xspace}
\newcommand{\ppbar}{\ensuremath{\Pp\overline{\Pp}}\xspace}
\newcommand{\PbPb}{\ensuremath{\text{PbPb}}\xspace}
\newcommand{\AuAu}{\ensuremath{\text{AuAu}}\xspace}

\newcommand{\sqrts}{\ensuremath{\sqrt{s}}\xspace}
\newcommand{\sqrtsnn}{\ensuremath{\sqrt{s_{NN}}}\xspace}

\newcommand{\mbinv} {\mbox{\ensuremath{\,\text{mb}^\text{$-$1}}}\xspace}
\newcommand{\mubinv} {\mbox{\ensuremath{\,\mu\text{b}^\text{$-$1}}}\xspace}

\newcommand{\ptt} {\ensuremath{p_{\mathrm{T}}}\xspace}

\providecommand{\CASCADE} {{\textsc{cascade}}\xspace}

\cmsNoteHeader{CMS Quarkonia Measurements} 
\title{Quarkonia Measurements by the CMS Experiment in pp and PbPb Collisions}

\address{{\it catherine.silvestre@cern.ch}\\LANL/UIC (until February 2011)\\LPSC, 53 rue des Martyrs, 38026 Grenoble Cedex, France}

\author{Catherine Silvestre, on the behalf of the CMS collaboration}

\date{\today}

\abstract{Quarkonia have been studied in different collision system and energy in order to understand the effects of the hot and dense medium created in heavy-ion collisions. CMS is well suited to measure quarkonia decays to muons given the muon identification and charged particle tracking capability. We report here prompt, non-prompt $\Jpsi$, and $\Upsilon$ production measured by the CMS experiment in \pp collisions at $\sqrt{s}=7$~TeV. In addition, the $\Jpsi$ and $\Upsilon$ production in \PbPb at $\sqrt{s_{NN}}=2.76$~TeV and \pp collisions at the same per nucleon energy are measured and compared. Prompt and non-prompt $\Jpsi$ contributions are separated for the first time in heavy-ion collisions, as is the ground from the excited states in the $\Upsilon$ family. Suppression in \PbPb at $\sqrt{s_{NN}}=2.76$~TeV is quantified for prompt $\Jpsi$, $B\rightarrow\Jpsi$, and $\Upsilon$(1S), as well as the relative suppression of $\Upsilon$(2S+3S) compared to $\Upsilon$(1S).}

\hypersetup{%
  pdfauthor={CMS Collaboration},%
  pdftitle={Quarkonium production in PbPb collisions},%
  pdfsubject={CMS},%
  pdfkeywords={CMS, physics, heavy ions, dimuons, quarkonia}}

\maketitle 

Quarkonia are especially relevant for studying the quark gluon plasma (QGP) since they are produced at early times and propagate through the medium mapping its evolution. In particular, $\Jpsi$ in heavy-ion collisions was suggested to be a promising probe as the deconfined medium should screen the two quarks leading to a suppression of its production\cite{Matsui:1986dk}. It has been studied at different energies and with different collision systems without yet giving a fully understood global picture~\cite{Baglin:1994ui,Alessandro:2006ju,Arnaldi:2007zz, Adare:2011yf}. 
Measuring the charmonium production at the LHC energies in \PbPb collisions will help constrain predictions, in particular those with a large recombination probability for prompt $\Jpsi$s. Indeed, the abundance of charm quarks in the medium could lead to a strong production enhancement at LHC energies~\cite{rapp}.
In addition to charmonium precision studies, the LHC center-of-mass energy allows copious $\Upsilon$ production in \PbPb collisions. Detailed measurements of bottomonia will help characterize the dense matter produced in heavy-ion collisions complementing the measurements accessible at RHIC energies. The full spectroscopy of quarkonium states has been suggested as a possible thermometer for the QGP~\cite{Miao:2010tk}. 

This paper first reviews CMS $\Jpsi$ and $\Upsilon$ cross-section measurements in \pp collisions at $\sqrt{s_{NN}}=7$~TeV, which allow precision studies of quarkonia production, as well as  at $\sqrt{s_{NN}}=2.76$~TeV, which will serve as a reference for the observation of hot nuclear effects in \PbPb at the same energy. CMS is able to distinguish non-prompt $\Jpsi$ from prompt $\Jpsi$ in both \pp and \PbPb collisions. The nuclear modification factor (\raa) of prompt, non-prompt $\Jpsi$, and $\Upsilon$(1S) in \PbPb is measured as a function of transverse momentum (\ptt), rapidity ($y$) and number of nucleons participating ($N_{part}$) in the collision~\cite{HIN-10-006}. Finally, the relative suppression of the excited states compared to the ground state is quantified~\cite{HIN-11-007}.

A detailed description of the CMS detector can be found in~\cite{ref:JINST}. Its central feature is a superconducting solenoid of 6~m internal diameter, providing a magnetic field of 3.8~T. Within the field volume are the silicon pixel and strip tracker, the crystal electromagnetic calorimeter, and the brass/scintillator hadron calorimeter. Muons are measured in gas-ionisation detectors embedded in the steel return yoke. In addition, CMS has extensive forward calorimetry, in particular two steel/quartz-fiber \v{C}erenkov hadron forward (HF) calorimeters, which cover the pseudorapidity range $2.9 < |\eta| < 5.2$.

In this paper, quarkonia are identified through their dimuon decay. The silicon pixel and strip tracker measures charged-particle trajectories for the  range $|\eta| < 2.5$. The tracker consists of 66M pixel and 10M strip detector channels, providing a vertex resolution of $\sim$\,15~$\mu$m in the transverse plane. Muons are detected for the $|\eta| < 2.4$ range, with detection planes based on three technologies: drift tubes, cathode strip chambers, and resistive plate chambers. Due to the strong magnetic field and the fine granularity of the silicon tracker, the muon transverse momentum measurement based on information from the silicon tracker alone has a resolution between 1 and 2\% for a typical muon in this analysis. CMS is therefore very well suited to measure dimuons. In pp, the resolution obtained measuring $\Jpsi$ in $|y|<0.5$ is 20~MeV/$c^2$, and 67~MeV/$c^2$ for the $\Upsilon$(1S) in $|\eta^\mu|<1$.

$\Jpsi$s can be classified into two types depending on whether they come from the primary vertex (prompt $\Jpsi$s) or are produced from decays of B mesons (non-prompt $\Jpsi$s). Prompt $\Jpsi$s group direct $\Jpsi$ production and $\Jpsi$s coming from the feed-down of higher states such as $\psi$' and $\chi_c$. Non-prompt $\Jpsi$s are produced at a distance $L_{xy}$ from the primary vertex and can therefore be separated from the prompt contribution if the resolution of the detector is good enough. This is done in CMS by reconstructing the $\mu^+\mu^-$ vertices and making a 2-dimensional simultaneous fit of the invariant mass distribution and the pseudo-proper decay length, $l_{\Jpsi}=L_{xy}\frac{m_{\Jpsi}}{p_T}$ (see~\cite{EPJC71:1515} for details).

\begin{figure}[h]
\begin{center}
\includegraphics[width=11pc,angle=90]{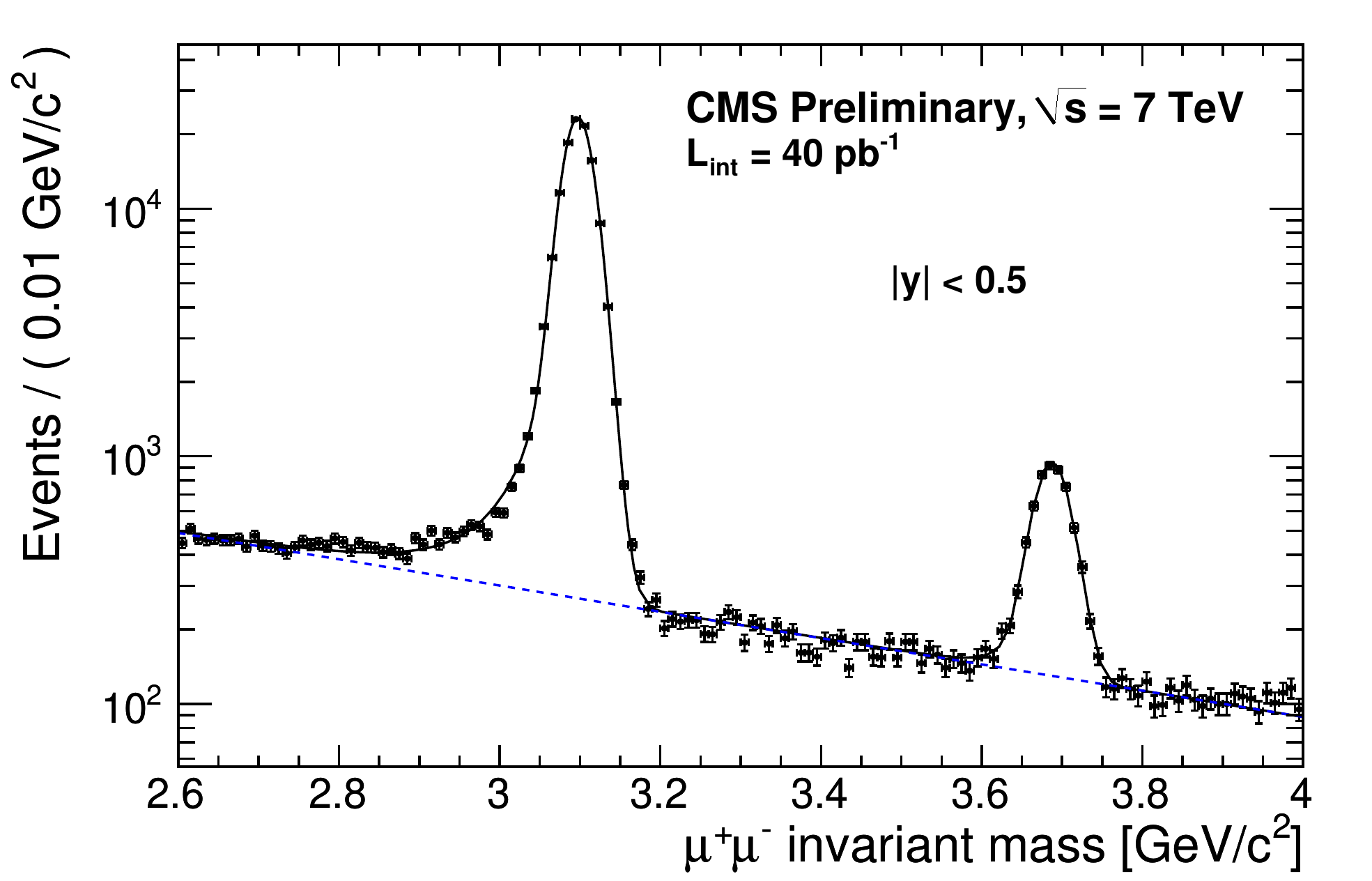}
\includegraphics[width=11.4pc,angle=90]{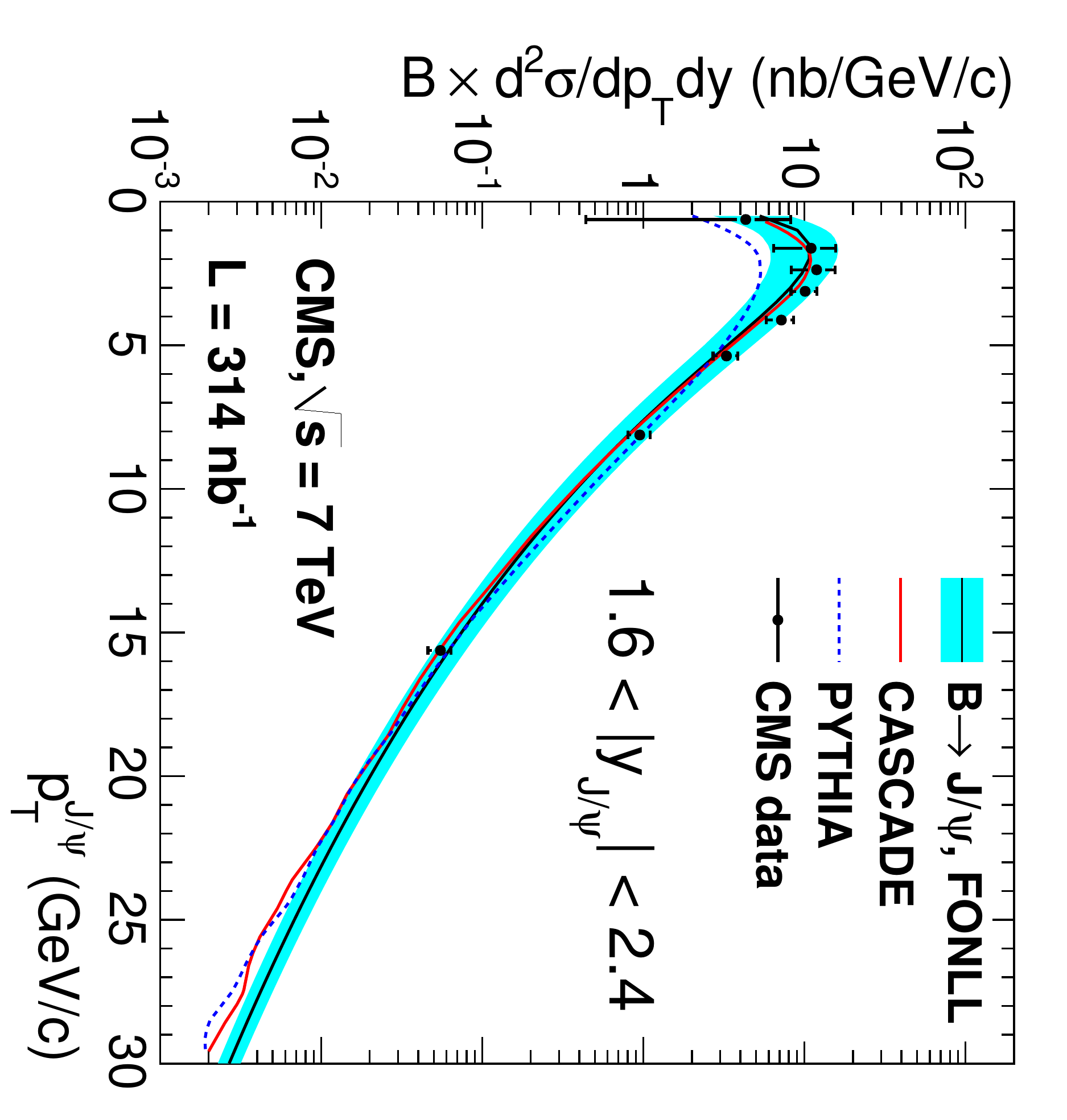}
\caption{\label{fig:ppJpsi}Left : invariant mass distribution showing the $\Jpsi$ and $\psi$' peaks for $|y|<0.5$.
 Right : $B\rightarrow\Jpsi$ cross-section as a function of $p_T$ for $1.6<|y|<2.4$ published in~\cite{EPJC71:1515} compared to theoretical predictions.}
\end{center}
\end{figure}

CMS has measured prompt and non-prompt $\Jpsi$ cross-sections as a function of $p_T$ in different rapidity bins~\cite{EPJC71:1515}. Fig.~\ref{fig:ppJpsi} (left) illustrates the inclusive invariant mass distributions measured with $\mathcal{L}_{\text{int}}=40$~nb$^{-1}$ 2010 \pp data at $\sqrt{s}=7$~TeV over $|y|<0.5$ with the clear $\Jpsi$ and the $\psi$'  peaks.  On Fig.~\ref{fig:ppJpsi} (right), the $B\rightarrow\Jpsi$ CMS cross-section measurement with $\mathcal{L}_{\text{int}}=314$~nb$^{-1}$ is overlaid with different predictions which reproduces well the data down to low $p_T$  in the forward region ($1.6<|y|<2.4$).

\begin{figure}[h]
\begin{center}
\includegraphics[width=18pc]{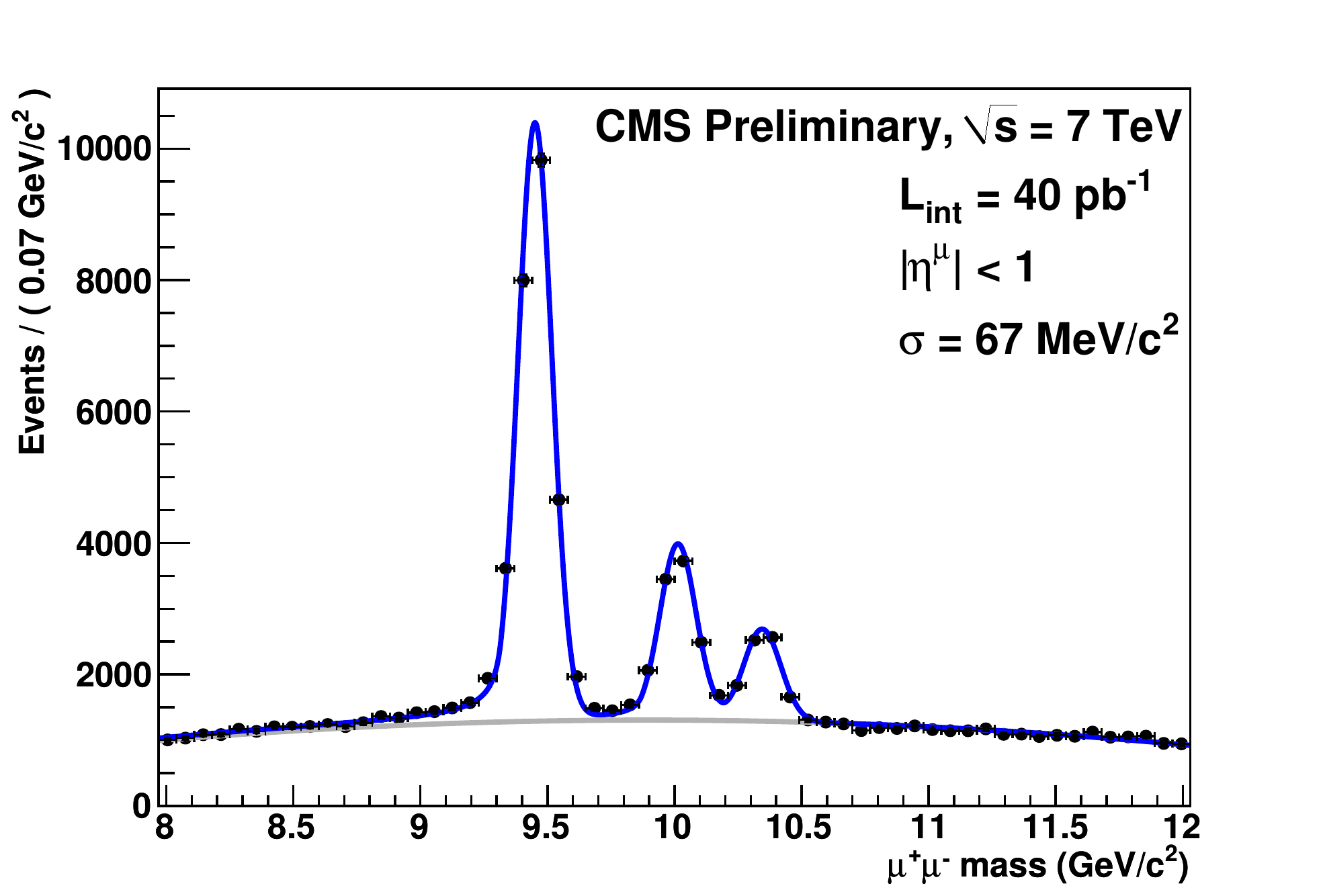}
\includegraphics[width=12.2pc]{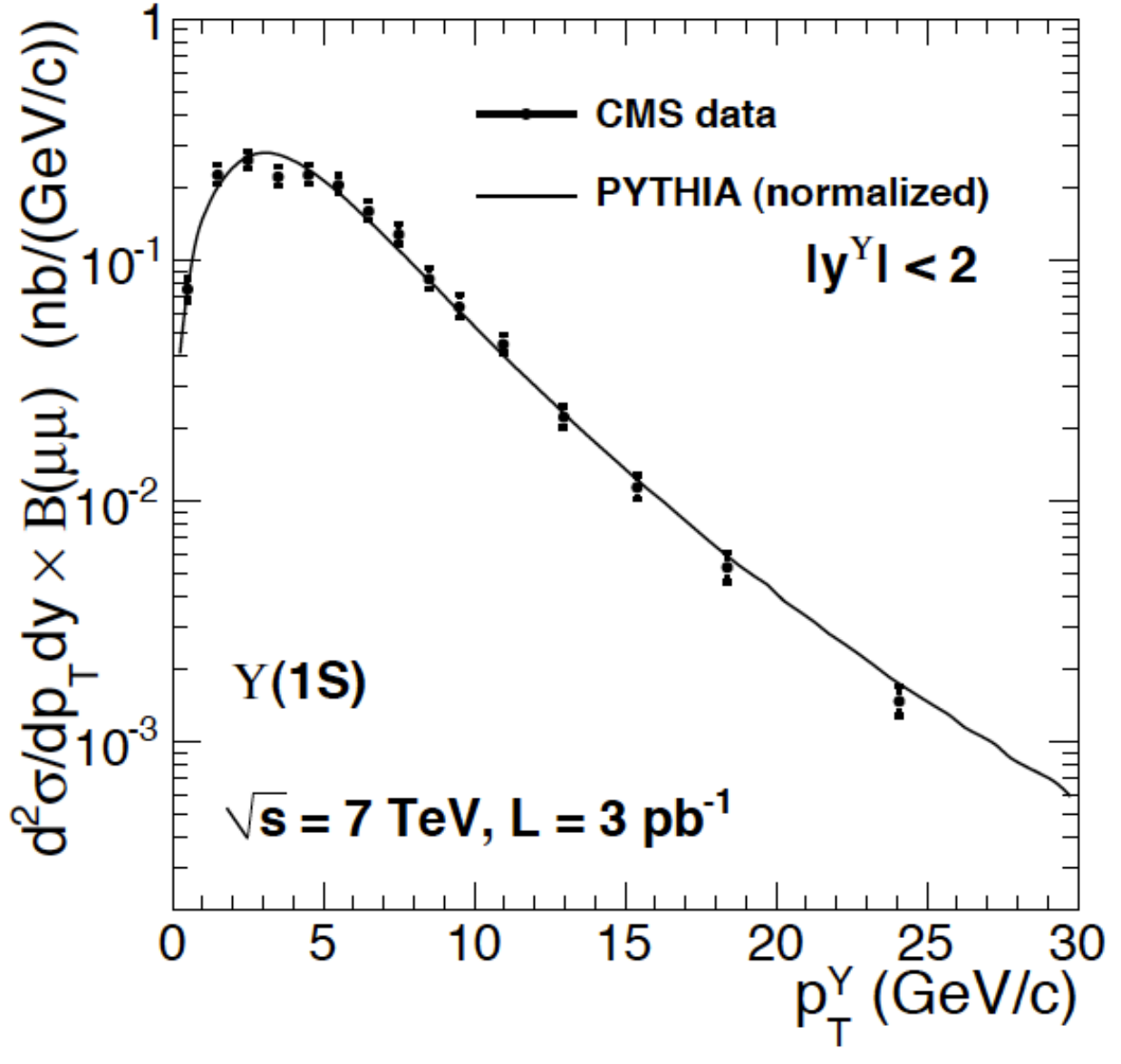}
\caption{\label{fig:ppupsilon}Left: $\Upsilon$ invariant mass distribution for $|\eta^\mu|<1$. Right : $\Upsilon$(1S) cross-section as a function of $p_T$ for $|y|<2$~\cite{ppUpsilon} compared to \textsc{Pythia}.}
\end{center}
\end{figure}

CMS is able to identify the three $\Upsilon$ states as shown on Fig.~\ref{fig:ppupsilon} (left) and has measured the $\Upsilon$(1S), $\Upsilon$(2S) and $\Upsilon$(3S) cross-sections as a function of $p_T$~\cite{ppUpsilon}. Fig.~\ref{fig:ppupsilon} (right) shows that the shape of the $\Upsilon$(1S) cross-section is well reproduced by \textsc{Pythia} but not the normalization which is overestimated by a factor two.

In March 2011, CMS recorded a little more than $\mathcal{L}_{\text{int}}=220$~nb$^{-1}$ \pp events at $\sqrt{s}=2.76$~TeV. This data is used as a reference for the \PbPb measurement. In November 2010, CMS recorded $\mathcal{L}_{\text{int}}=7.28$~$\mu b^{-1}$ of \PbPb events, leading to about the same amount of quarkonia statistics as the reference \pp run at 2.76~TeV. Both data sets have been analyzed following similar conditions~\cite{HIN-10-006,HIN-11-007}: (1) events are selected by the CMS two-level trigger keeping any dimuon activity in the muon chambers, (2) offline muon reconstruction is seeded with $\simeq 99$\% efficient tracks in the muon detectors, which are then matched to tracks reconstructed in the silicon tracker by means of an algorithm optimized for the heavy-ion environment~\cite{Roland:2006kz,D'Enterria:2007xr}, (3) the same analysis procedure is followed for the offline selection with very loose criteria.  Signal extraction is based on the procedures in CMS 7~TeV publications for the signal extraction~\cite{EPJC71:1515,ppUpsilon}.

Using both data sets, the production measured in \PbPb collisions is compared to expectations from an independent superposition of nucleon-nucleon collisions typically expressed in terms of the nuclear modification factor:
\begin{equation}
  \raa = \frac{\mathcal{L}_{\pp}}{\taa N_{\text{MB}}}\frac{N_{\PbPb} (\QQbar)}{N_{\pp} (\QQbar)}\cdot \frac{\varepsilon_{\pp}}{\varepsilon_{\PbPb}}.
\end{equation}
Here $\taa$ is the nuclear overlap function\footnote{Ratio of the number of binary nucleon-nucleon collisions $N_{coll}$ calculated from a Glauber model of the nuclear collision geometry~\cite{centrality,PhysRevC.77.014906} and the inelastic nucleon-nucleon cross section $\sigma^{NN}_{inel} = (64\pm 5)$~mb at $\sqrt{s} = 2.76$~TeV~\cite{PDG}}, $\mathcal{L}_{\pp}$ is the \pp luminosity, $N_{\text{MB}}$ is the measured number of equivalent minimum bias events in \PbPb , $\frac{N_{\PbPb} (\QQbar)}{N_{\pp} (\QQbar)}$ is the raw yield ratio, and $\frac{\varepsilon_{\pp}}{\varepsilon_{\PbPb}}$ the multiplicity dependent
fraction of the efficiency ($\frac{\varepsilon_{\pp}}{\varepsilon_{\PbPb}} \sim 1.17$ for
the most central bin).

\begin{figure}[h]
\begin{center}
\includegraphics[width=14pc]{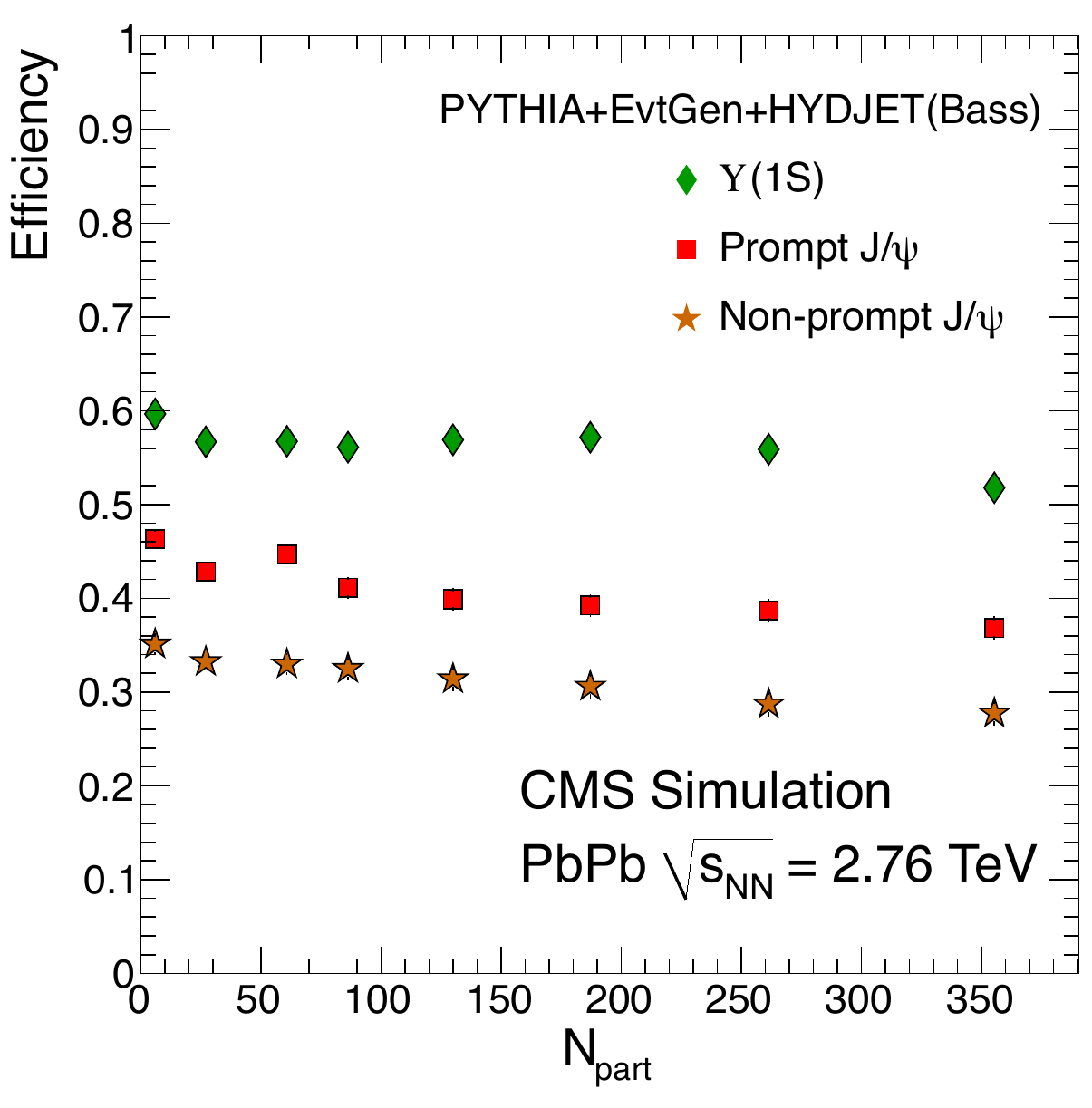}
\includegraphics[width=14.8pc]{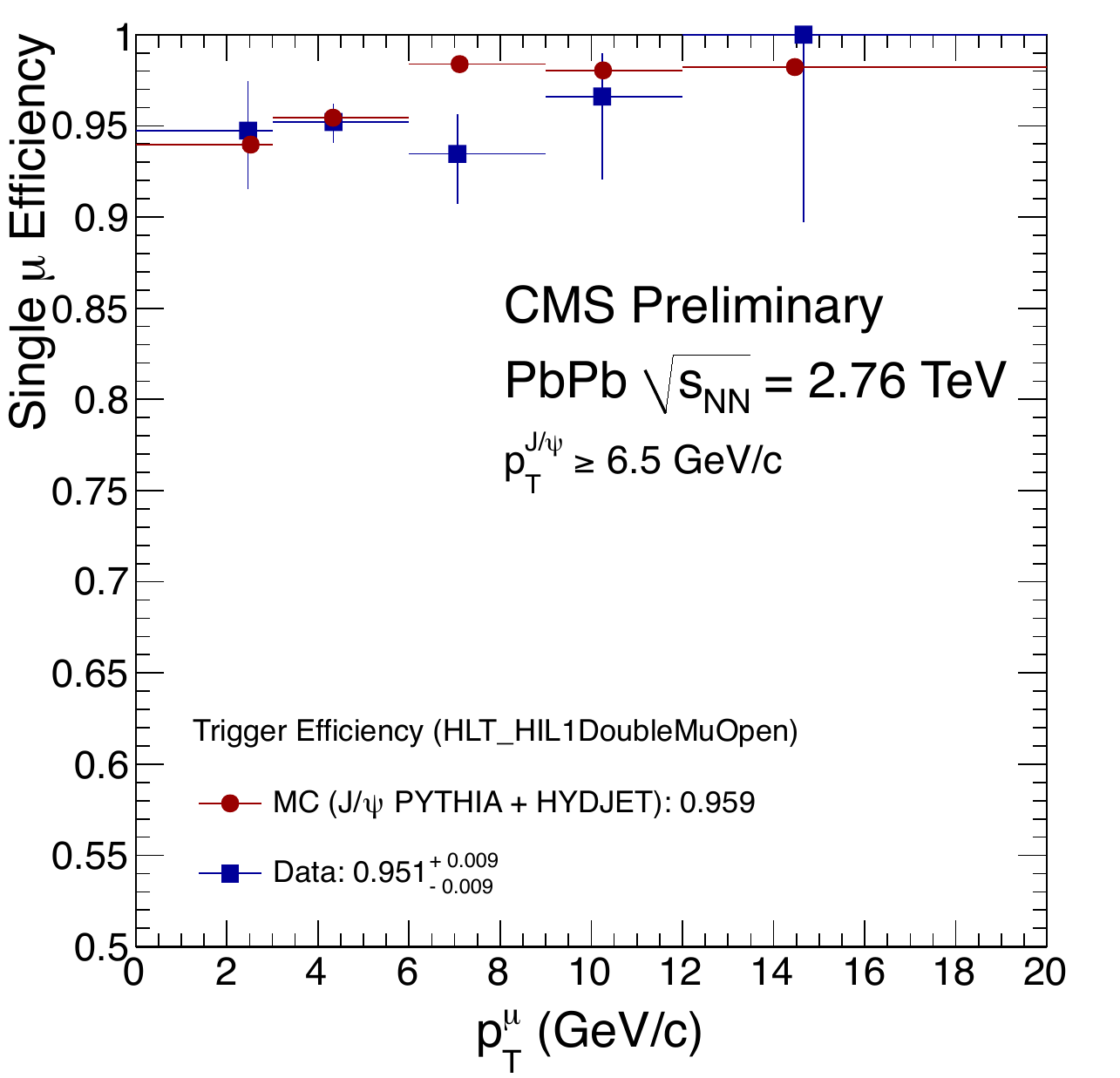}
\caption{\label{fig:hiEff}Left : $\Upsilon$ (diamonds), prompt (squares) and non-prompt (stars) $\Jpsi$ MC efficiency as a function of $p_T$ in \PbPb collisions. Right : single muon efficiency for the double muon trigger used as a function on $p_T$ in MC (circles) and data (squares).}
\end{center}
\end{figure}

Trigger, reconstruction and selection efficiencies of muon pairs are estimated using quarkonia \textsc{Pythia} signal embedded in heavy-ion \PbPb events generated by \textsc{Hydjet}~\cite{Lokhtin:2005px}. These events were processed through the trigger emulation and event reconstruction chain. The final efficiency corrections correspond to the fraction of reconstructed signal passing all the analysis selections with respect to the generated signal. Fig.~\ref{fig:hiEff} (left) shows this dimuon efficiency as a function of $N_{part}$ for $\Upsilon$ (diamonds), prompt (circles) and non-prompt (stars) $\Jpsi$. The individual components of the MC efficiency are cross checked using a {\it tag-and-probe} technique applied to data~\cite{EPJC71:1515}. The method consists of
fitting the \Jpsi candidates, with and without applying the probed selection on one of the muons. For example, the dimuon trigger efficiency is probed by testing the trigger response from a single-muon triggered sample. Fig.~\ref{fig:hiEff} (right) shows that the single muon efficiency for the dimuon trigger used in the analysis as a function of $p_T$ measured in data ($95.1\pm0.9$)\% (squares) is compatible with MC (circles).

\begin{figure}[h]
\begin{center}
\includegraphics[width=14pc]{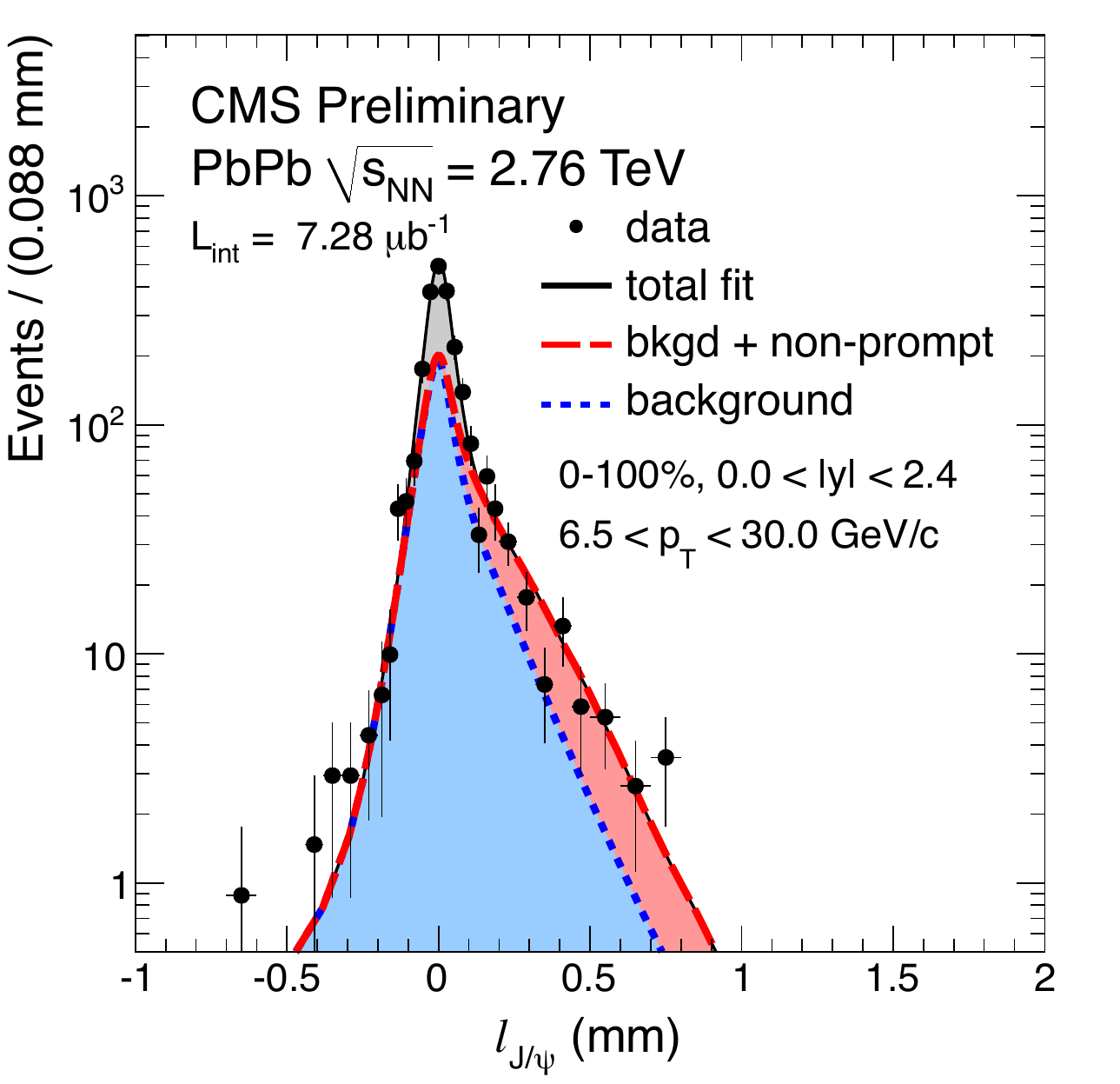}
\includegraphics[width=14pc]{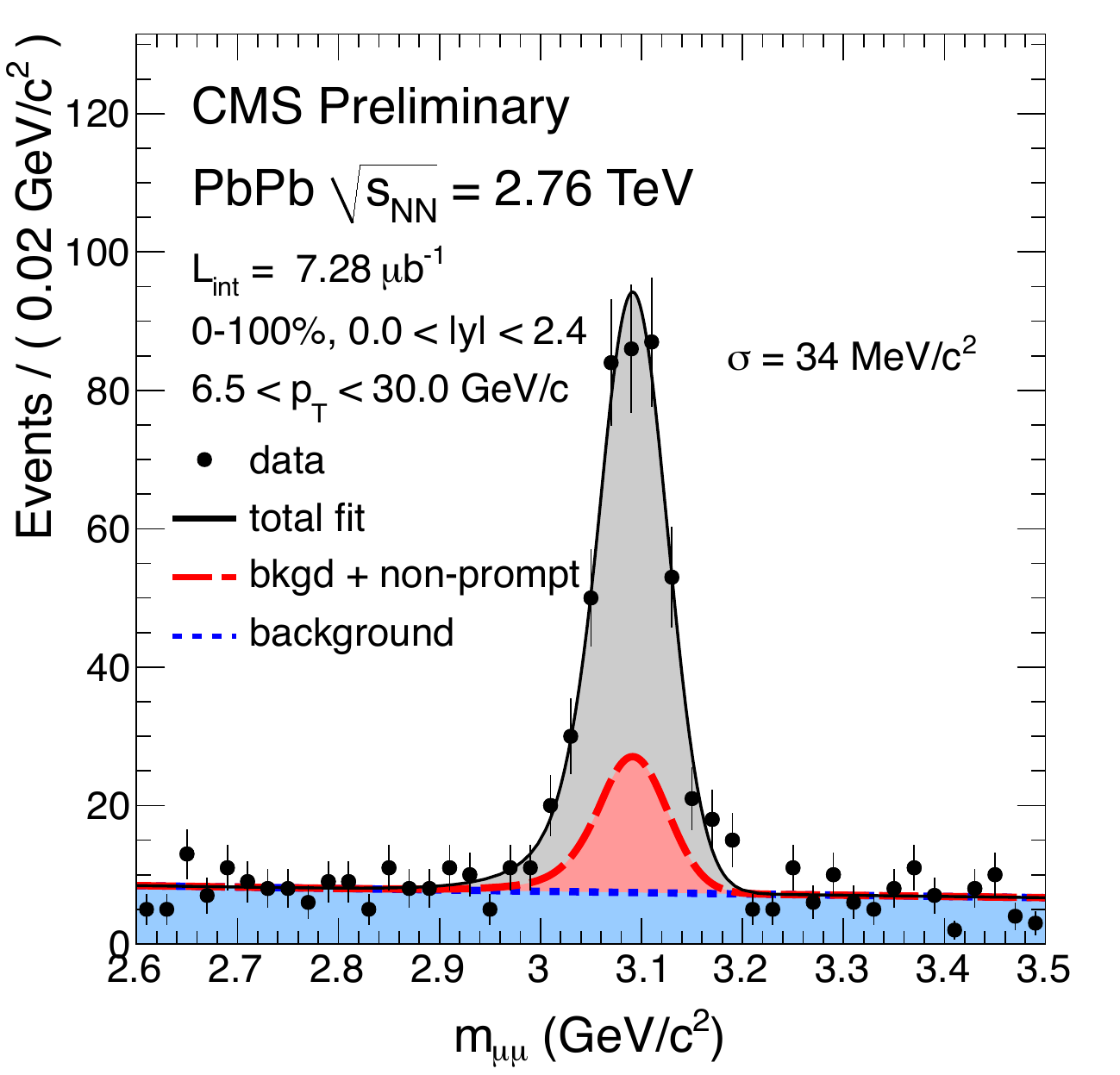}
\caption{\label{fig:PbPbNonpropmtFit}Pseudo-proper decay length (left) and invariant mass distribution (right) for $\Jpsi$ with $p_T>6.5$~GeV/$c$ in \PbPb collisions.The black dots are the data, the dotted line filled in blue the background, the dashed line filled in red the background and non-prompt contribution, and the black straight line the total fit.}
\end{center}
\end{figure}

The CMS detector performs very well in heavy-ion environment such that the good momentum resolution can be used to separate non-prompt from prompt $\Jpsi$ as in pp, making use of the distance between the non-prompt vertex and the primary vertex. An example of the 2-dimensional fit in \PbPb collisions is shown on Fig.~\ref{fig:PbPbNonpropmtFit} for $\Jpsi$ with $\ptt>6.5$~GeV/$c$. For more details see~\cite{HIN-10-006,Torsten,Mihee}.

\begin{figure}[h]
\begin{center}
\includegraphics[width=14pc]{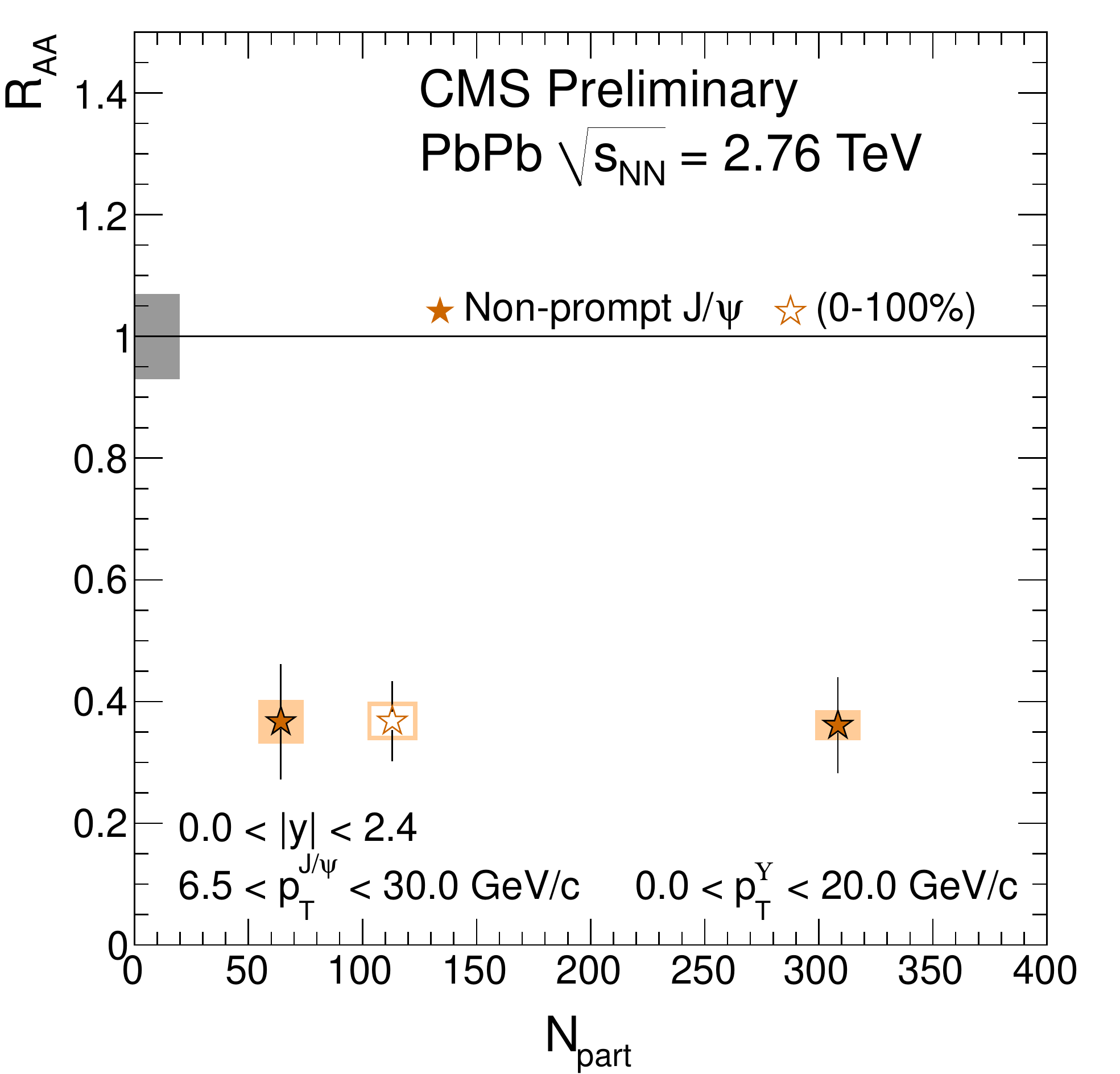}
\includegraphics[width=14pc,angle=90]{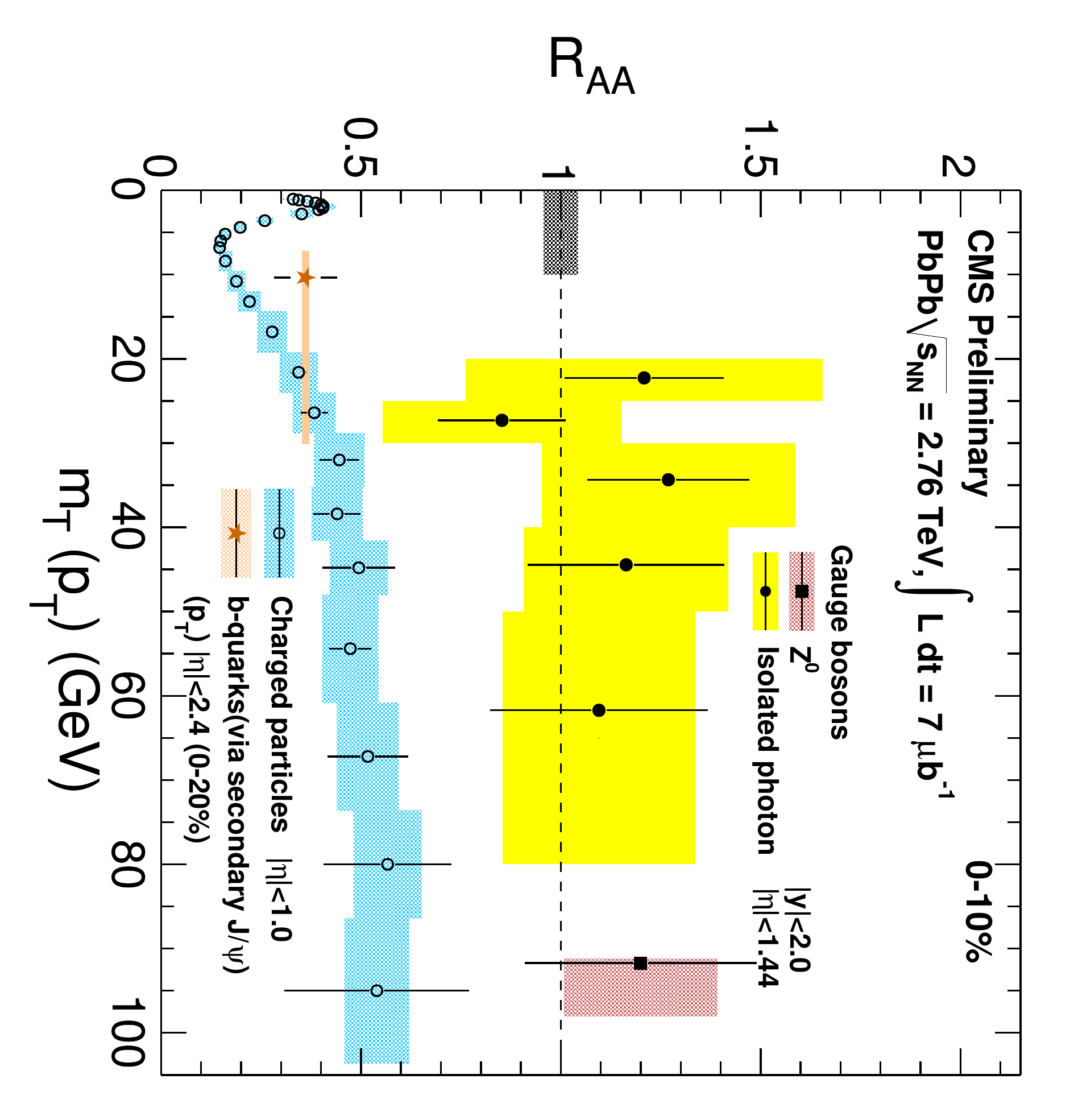}
\caption{\label{fig:nonPromptRaa}Left: Non-prompt $\Jpsi$ \raa in three centrality bin 0--20\% and 20--100\% in closed symbols and 0--100\% in open symbols. Right: \raa vs. $m_T$ for Z (squares), isolated photons (closed circles) and charged hadrons (open circles) compared to the secondary $\Jpsi$ measurement as a function of $\ptt$ (stars).}
\end{center}
\end{figure}

For the first time, secondary $\Jpsi$ \raa is measured in heavy-ion collisions. Fig.~\ref{fig:nonPromptRaa} (left) illustrates B-meson suppression through their $\Jpsi$ decays through the \raa as a function of $N_{part}$: $\raa =
0.36\pm0.08(\text{stat})\pm0.03(\text{syst})$ in the 20\% most central collisions. This could be a hint of b-quark energy loss. The level of suppression is of the same order of magnitude as charged hadrons as observed on Fig.~\ref{fig:nonPromptRaa} (right) where the non-prompt $\Jpsi$ \raa is plotted as a function of \ptt for 0--20\%  while the bosons and charged hadrons are presented as a function of the transverse mass for 0--10\%~\cite{YenJie}. 

\begin{figure}[h]
\begin{center}
\includegraphics[width=11pc]{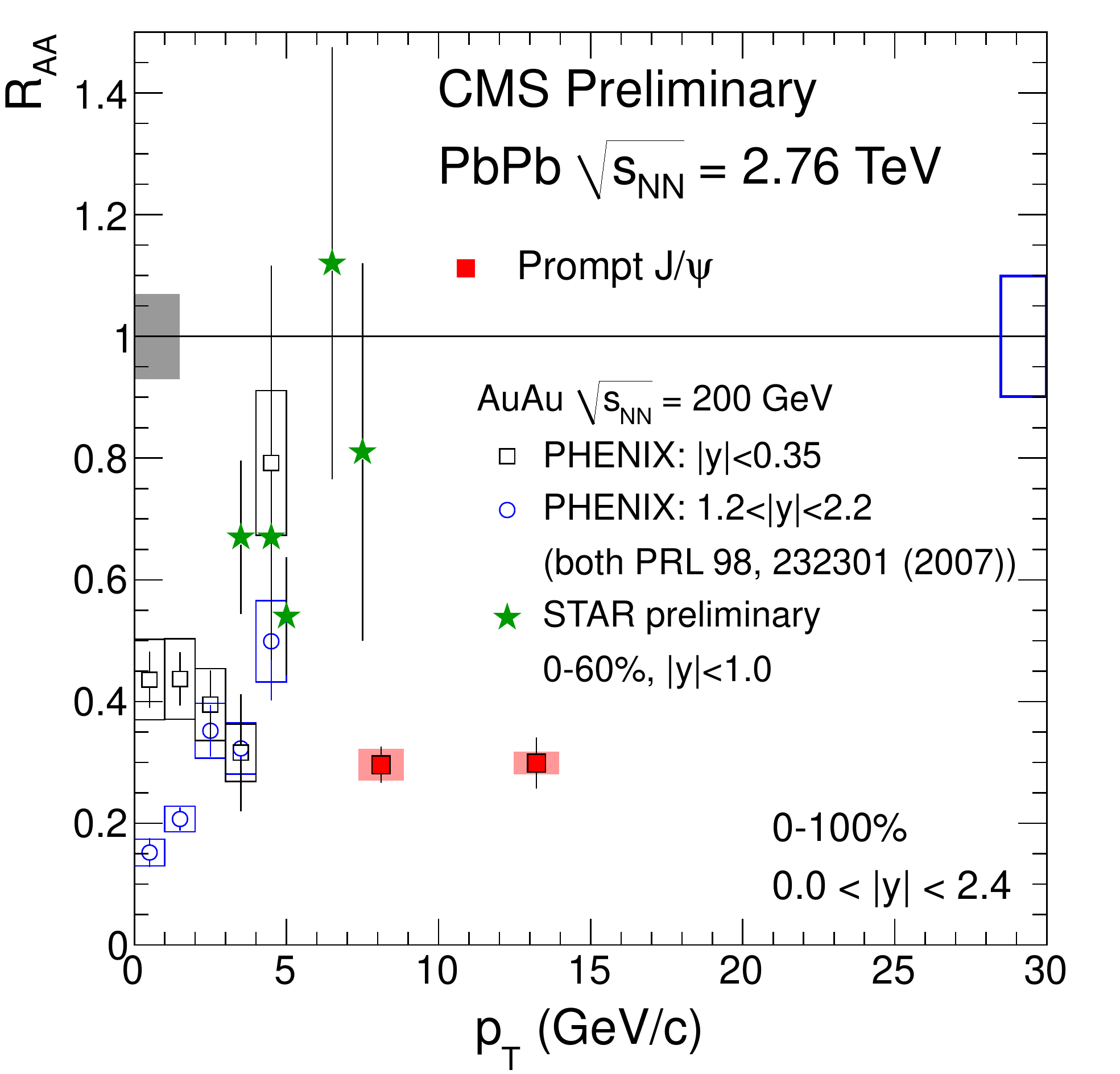}
\includegraphics[width=11pc]{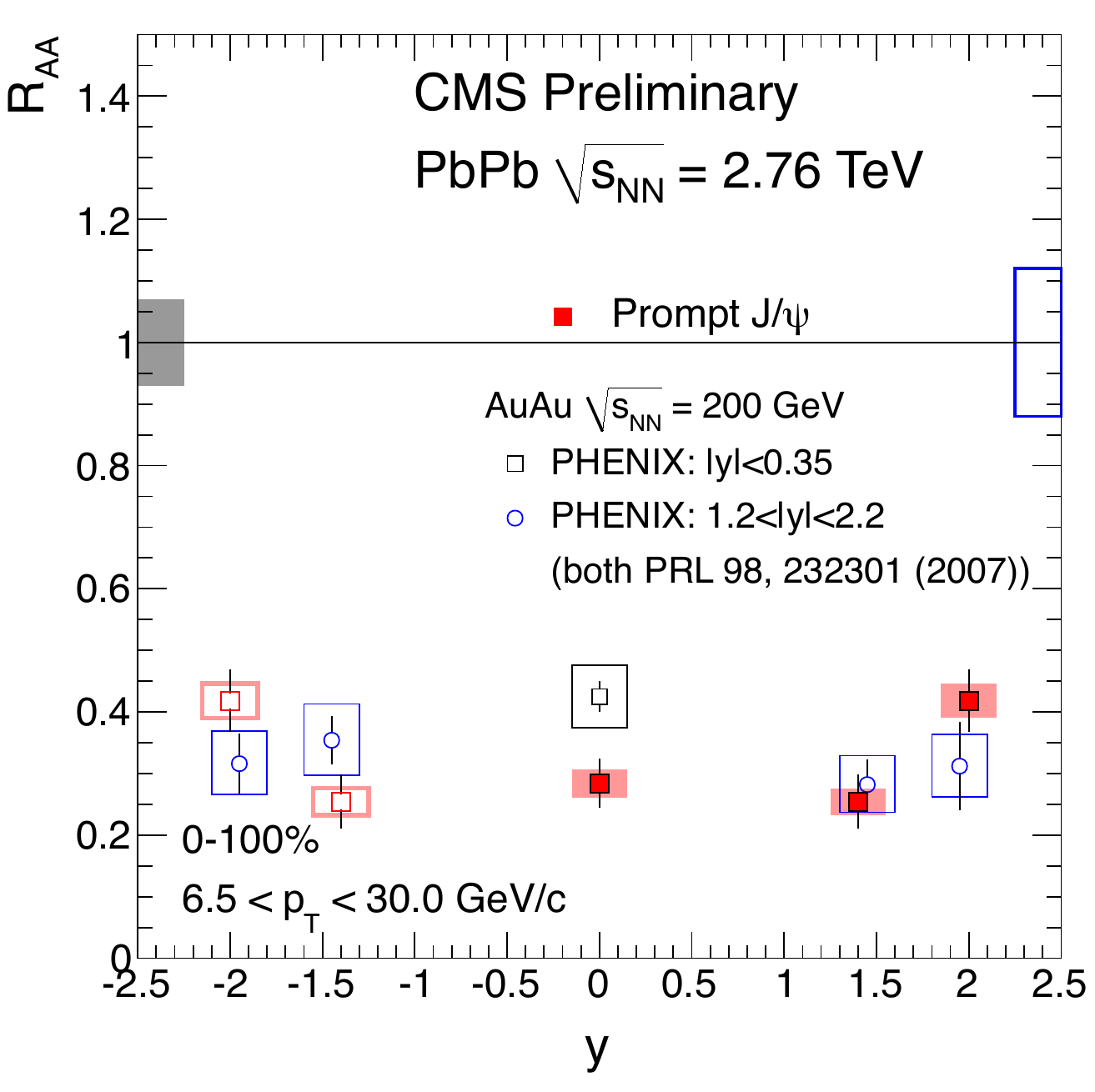}
\includegraphics[width=11.5pc]{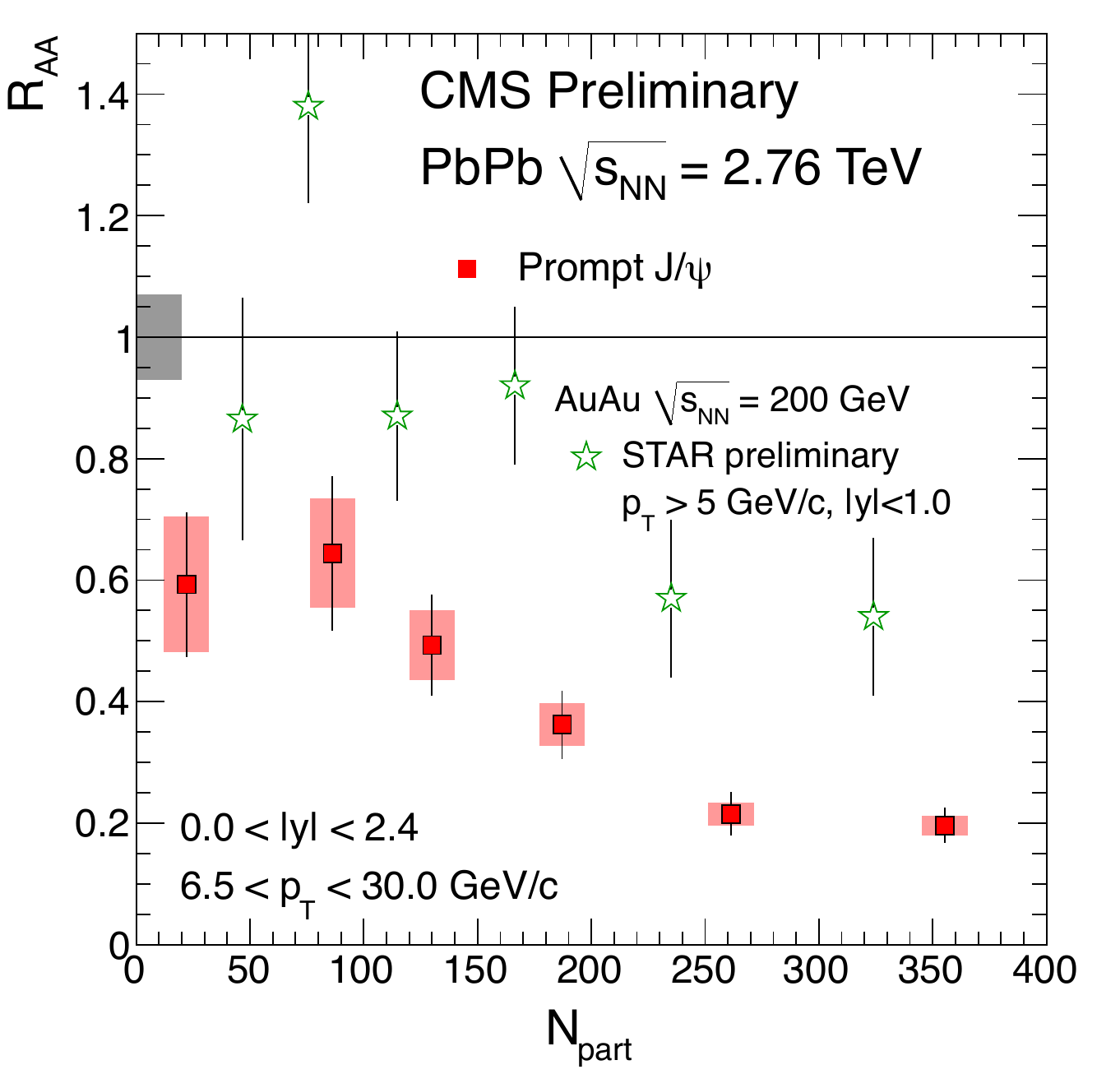}
\caption{\label{fig:propmtRaa}CMS prompt $\Jpsi$ \raa measurement (filled squares) compared to PHENIX mid (open squares) and forward (open circles) rapidity measurement and STAR higher $\ptt$ measurement (stars) as a function of $\ptt$ (left), $y$ (center), and $N_{part}$ (right).}
\end{center}
\end{figure}

Fig.~\ref{fig:propmtRaa} shows the prompt $\Jpsi$ \raa (filled squares) as a function of \ptt, $y$ and $N_{part}$. A factor three suppression is observed for the two \ptt bin. CMS points are compared to measurements at $\sqrt{s_{NN}}=200$~GeV from PHENIX~\cite{PHENIX} at mid- (open squares) and forward (open circles) rapidity for lower $\ptt$s, and from STAR~\cite{STAR} up to $\ptt=8$~GeV/$c$\footnote{PHENIX and STAR measurements are inclusive measurements but the contamination from secondary $\Jpsi$ is expected to be small at RHIC energies.}. The tendency of high $\ptt$ $\Jpsi$'s to survive at RHIC is not seen at the LHC. Furthermore, CMS measures less suppression at forward rapidity for high $\ptt$ $\Jpsi$. One should remember that the $x$ probed with $\langle \ptt^{\Jpsi}\rangle=10$~GeV/$c$ by CMS over $|y|<2.4$ are $x_1\sim 0.02$ and $x_2\sim 5\cdot 10^{-4}$ . Therefore, anti-shadowing could play a role in the suppression observed and could contribute to seeing an opposite trend than PHENIX as a function of $y$, or an increase of the \raa when going to low $\ptt$ and more forward regions as for ALICE measurements~\cite{alice}. For $p_T>3$~GeV/$c$ and $1.6<y<2.4$, CMS measures $\raa=0.39\pm0.06\text{(stat.)}\pm0.03\text{(syst.)}$. Finally, in the 10\% most central collisions, CMS observes a factor five
suppression much greater than measured by STAR.

\begin{figure}[h]
\begin{center}
\includegraphics[width=14pc]{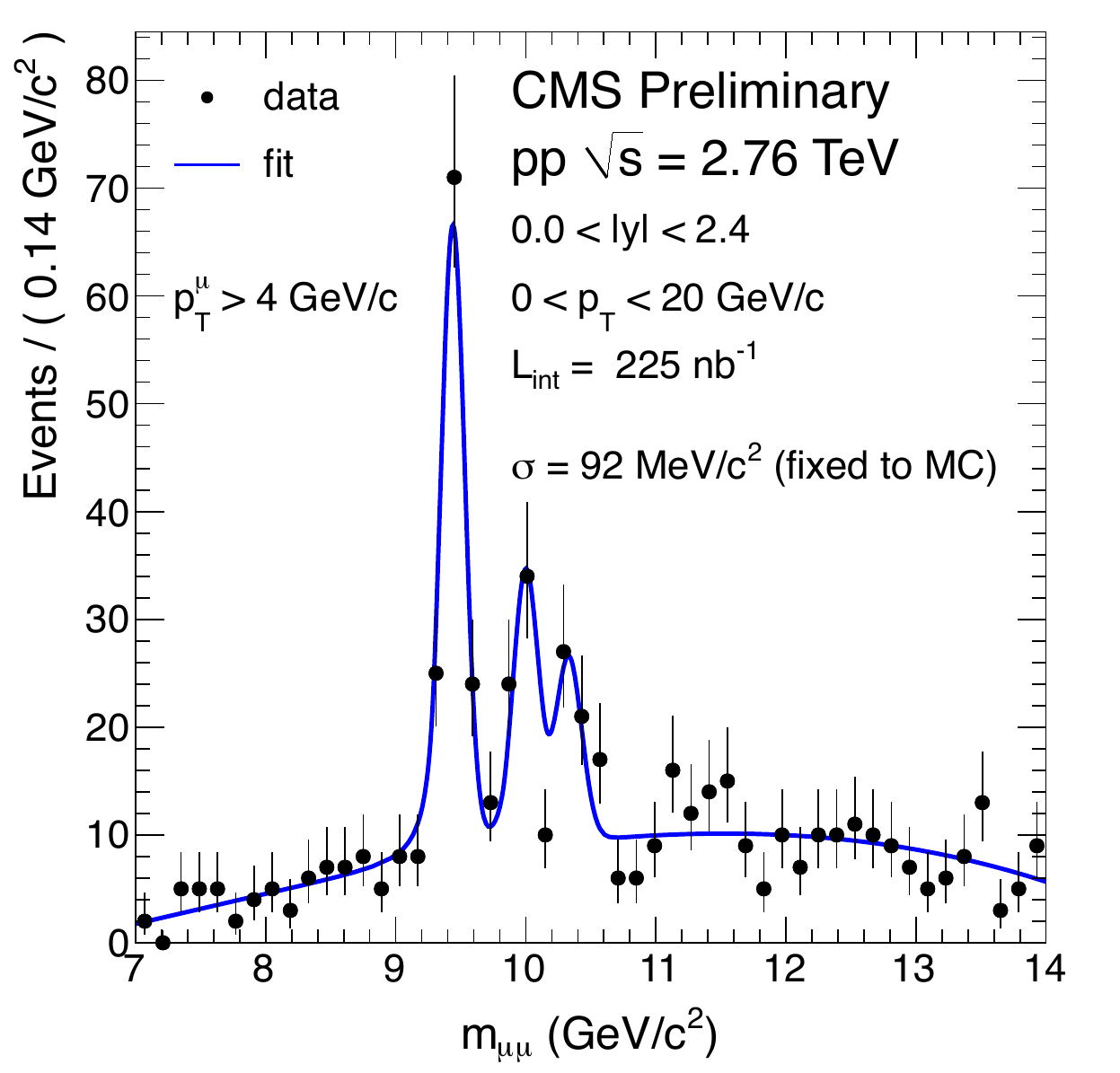}
\includegraphics[width=14pc]{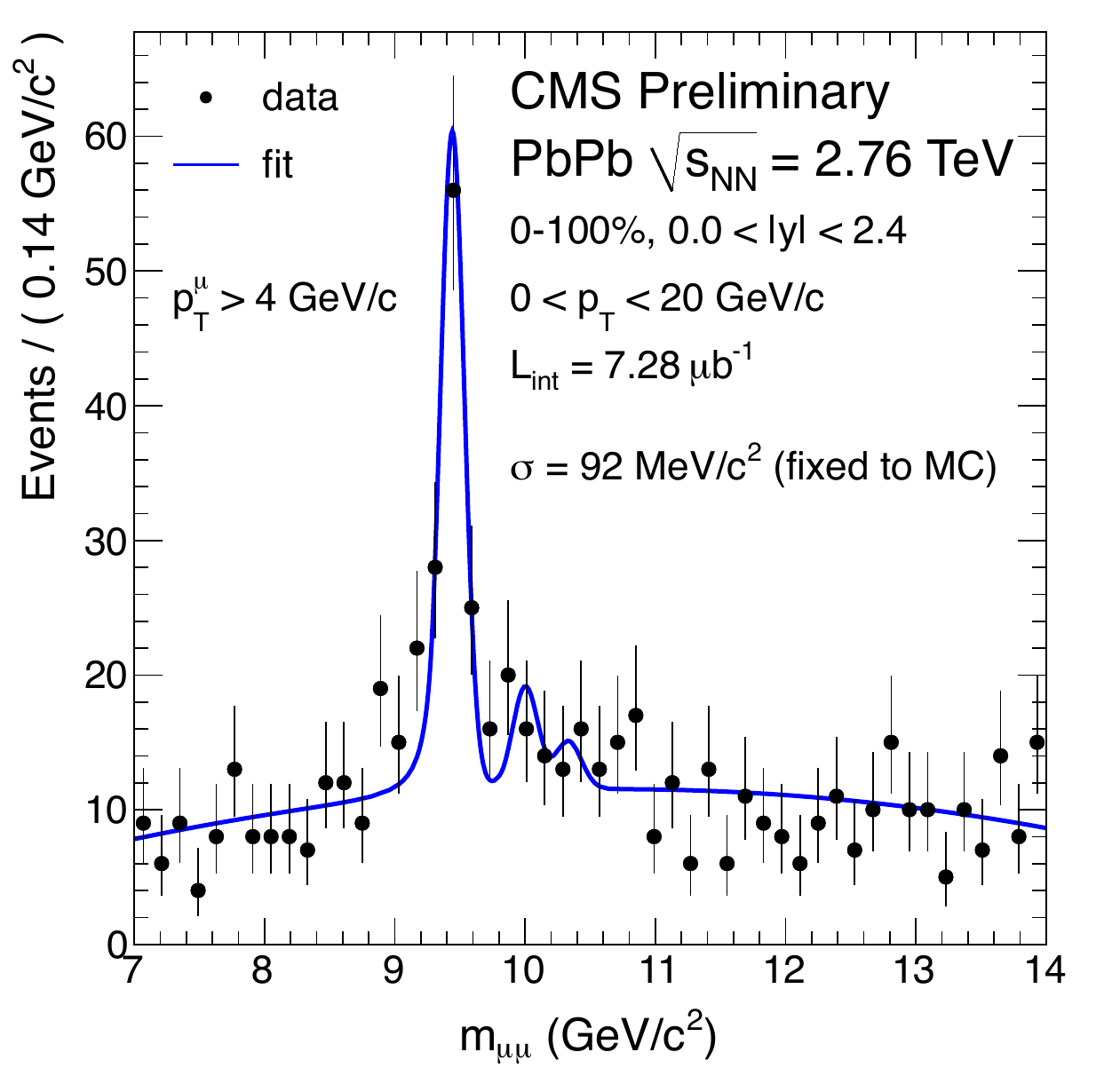}
\caption{\label{fig:upsExcited}Invariant mass distribution measured in \pp (left) and \PbPb (right) collisions at 2.76~TeV for $\ptt^\mu>4$~GeV/$c$.}
\end{center}
\end{figure}

CMS is able to disentangle the $\Upsilon$(1S) contribution from the higher states in \PbPb as in \pp collisions. Fig.~\ref{fig:upsExcited} compares the $\Upsilon$ invariant mass distribution at $\sqrt{s}=2.76$~TeV in \pp (left) and \PbPb (right) collisions, for $\ptt^\mu>4$~GeV/$c$. The higher state contribution relative to the ground state is strikingly smaller in \PbPb collisions. In order to quantify this suppression, an extended unbinned maximum likelihood simultaneous fit to the \pp and \PbPb mass spectra is performed, following the method described in~\cite{HIN-11-007}, using the parameters detailed in~\cite{Zhen}. The ratio of $\Upsilon(2S+3S)/\Upsilon(1S)$ in \PbPb and \pp benefits from an almost complete cancellation of possible acceptance and/or efficiency differences among the reconstructed resonances. The double ratio obtained is
\begin{equation}
  \frac{\Upsilon(2S+3S)/\Upsilon(1S)|_{\rm \PbPb}}{\Upsilon(2S+3S)/\Upsilon(1S)|_{\pp}}
 = 0.31 _{-0.15}^{+0.19} \; ({\rm stat.}) \pm 0.03 \; ({\rm syst.}),
\label{eq:double}
\end{equation}
where the systematic uncertainty (9\%) arises from varying the lineshape in the simultaneous fit, thus taking into account partial cancellations of systematic effects. Finally, using an ensemble of one million pseudo-experiments generated with the signal lineshape obtained from the \pp\ data, Fig.~\ref{fig:upsExcited} (left), the background lineshapes from both data sets, and a double ratio (Eq.~\ref{eq:double}) equal to unity within statistical and systematic uncertainties (absence of a suppression), the probability of finding the measured value of 0.31 or a downward fluctuation is estimated to be 0.9\%., corresponding to 2.4 sigma in a one-tailed integral of a Gaussian distribution.

\begin{figure}[h]
\begin{center}
\includegraphics[width=12pc]{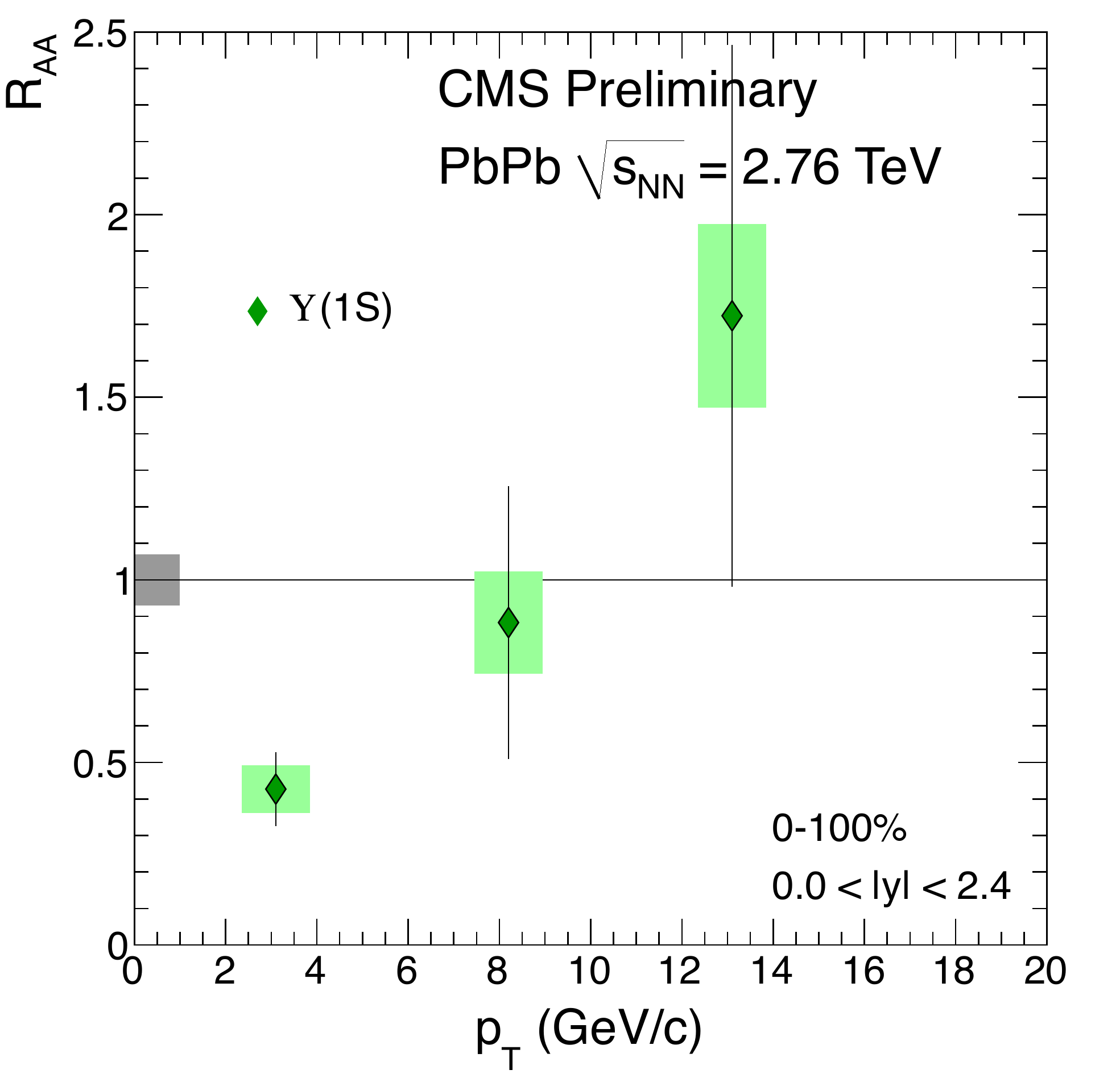}
\includegraphics[width=12pc]{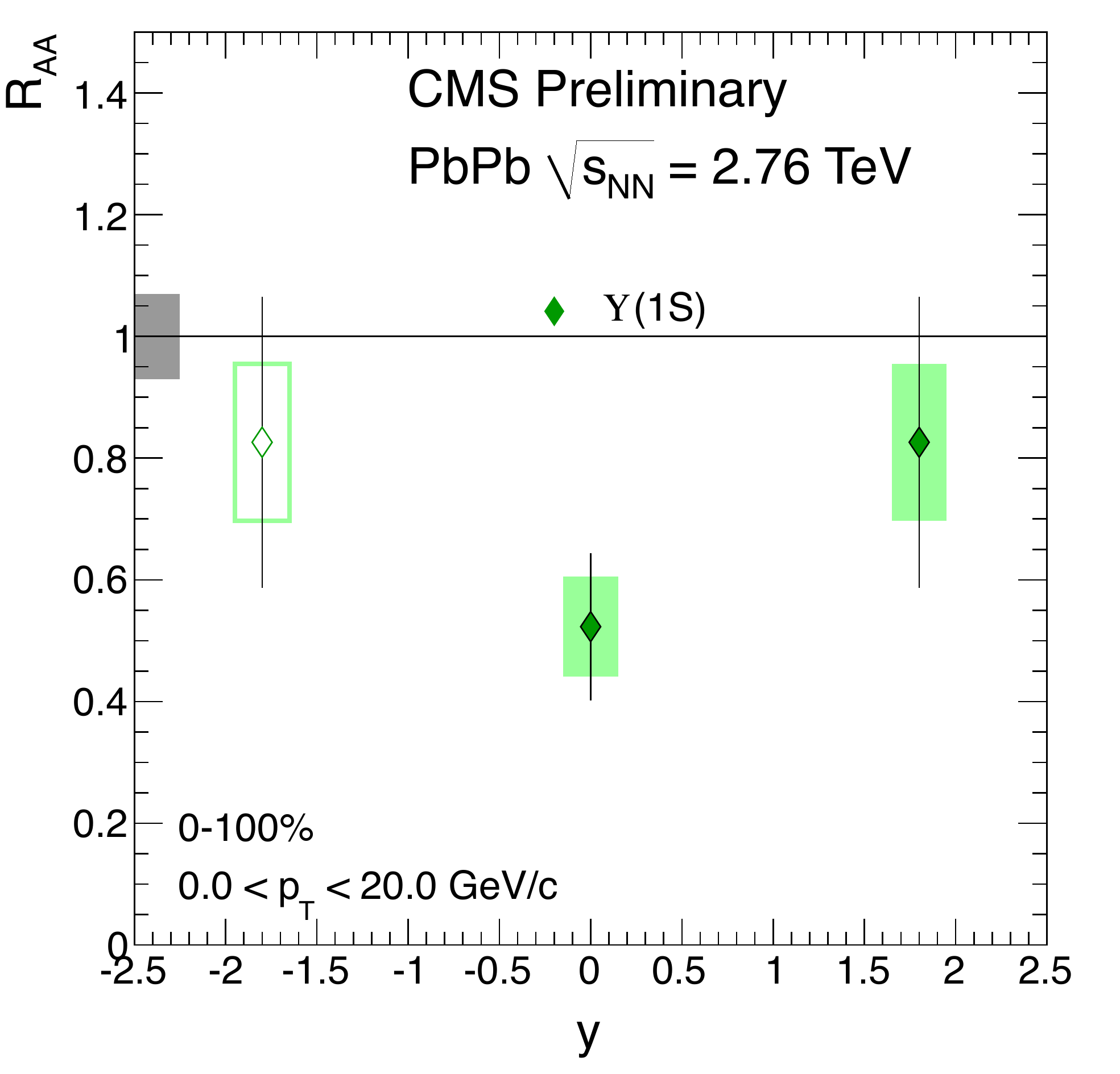}
\includegraphics[width=12pc]{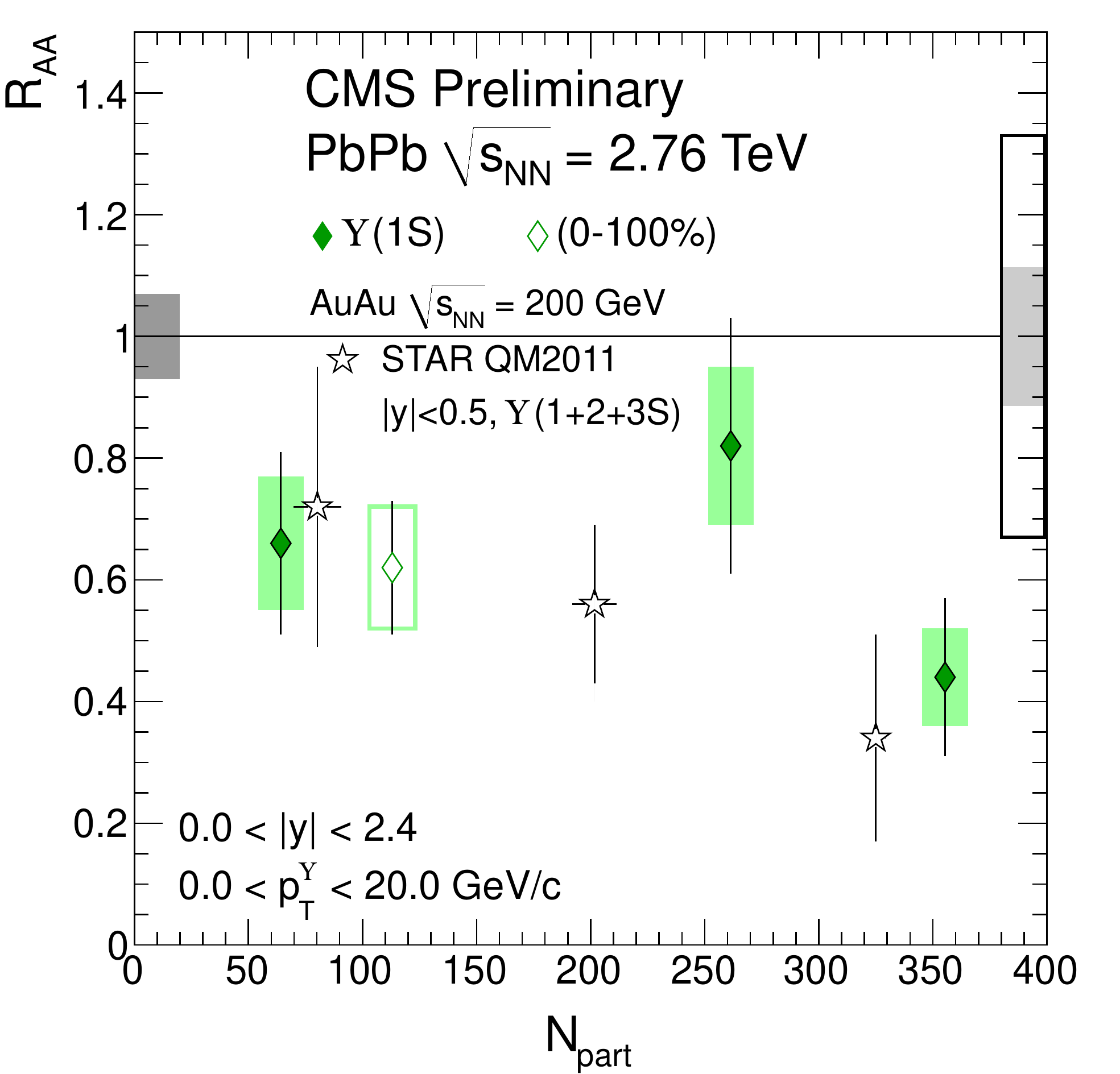}
\caption{\label{fig:upsRaa}$\Upsilon$(1S) \raa as a function of \ptt (left), $y$ (middle), and $N_{part}$ (right), compared to STAR inclusive preliminary measurement (stars) for the latter.}
\end{center}
\end{figure}

The $\Upsilon$(1S) suppression has a been studied as a function of $\ptt$, $y$ and centrality as shown on Fig.~\ref{fig:upsRaa}. A suppression by a factor $\sim 2.3$ is observed for low $\ptt$. This seems to disappear for $\ptt > 6.5\GeVc$. The rapidity dependence indicates a slightly smaller suppression at forward rapidity. In both cases however, the statistical uncertainties are too large for any strong conclusions. In addition, $\Upsilon$(1S) are suppressed by a factor two in 0--10\% central collisions. The CMS measurement is compared to STAR $\Upsilon$(1+2+3S) preliminary result in AuAu collisions at $\sqrt{s_{NN}}=200$~GeV~\cite{starUps} showing a suppression of the same order of magnitude but with large uncertainty.

In summary, this paper first presented highlights of CMS quarkonia measurements in \pp collisions at $\sqrt{s}=7$~GeV recorded in 2010. The high statistics accumulated at the LHC allows to perform details studies that further constrain production mechanisms. CMS has performed the measurements of the prompt and non-prompt \Jpsi, as well as of the $\Upsilon\text{(1S)}$ and $\Upsilon\text{(2S+3S)}$ mesons via their
decay into \mumu pairs in \PbPb and \pp collisions at 2.76~TeV. Prompt \Jpsi has been separated from non-prompt \Jpsi for the
first time in heavy-ion collisions. A strong suppression of prompt $\Jpsi$ with $p_T>6.5$~GeV/$c$ is measured in central collisions, and already in peripheral collisions, showing a clear dependence with centrality. Non-prompt \Jpsi, though strongly suppressed, show no strong centrality dependence within
uncertainties. This is the first hint of b-quark energy loss in the hot medium. Furthermore, $\Upsilon$(1S) are suppressed by 40\% in the 20\% most central collisions. The comparison of the ratios of ${\ensuremath{\Upsilon\text{(nS)}}\xspace}$-states in \pp and \PbPb
collisions, taken at the same center-of-mass energy, is consistent with the partial disappearance of the higher states with respect to the ground state in the \PbPb collisions. Those two observations could indicate that the $\Upsilon$(1S) suppression is due to the melting of the excited states only in \PbPb collisions. Measuring the amount of suppression caused by shadowing through pA collisions together with more precise measurements will be crucial for interpreting what the melting is due to.

\end{document}